# Title: Interfacial-Redox-Induced Tuning of Superconductivity in YBa$_2$Cu$_3$O$_{7-\delta}$

**Authors:**

*Peyton D. Murray,[1][‡] Dustin A. Gilbert,[3,4][‡] Alexander J. Grutter,[3] Brian J. Kirby,[3] David Hernandez-Maldonado,[5] Maria Varela,[5] Zachary E. Brubaker,[1,6] W.L.N.C. Liyanage,[4] Rajesh V. Chopdekar,[7,8] Valentin Taufour,[1] Rena J. Zieve,[1] Jason R. Jeffries,[9] Elke Arenholz,[8] Yayoi Takamura,[7] Julie A. Borchers,[3] Kai Liu[1,2]\**

**Author Addresses:**

[1]Physics Department, University of California, Davis, CA 95616, USA.

[2]Physics Department, Georgetown University, Washington, DC 20057, USA.

[3]NIST Center for Neutron Research, National Institute of Standards and Technology, Gaithersburg, MD 20899, USA.

[4]Department of Materials Science and Engineering, University of Tennessee, Knoxville, TN 37996, USA

[5]Dept. de Física de Materiales & Instituto Pluridisciplinar, Universidad Complutense de Madrid, Madrid 28040, Spain.

[6]Materials Science Division, Lawrence Livermore National Laboratory, Livermore, CA 94550, USA.

[7]Department of Materials Science and Engineering, University of California, Davis, CA 95616, USA.

[8]Advanced Light Source, Lawrence Berkeley National Laboratory, Berkeley, CA 94720, USA.

[9]Physics Division, Lawrence Livermore National Laboratory, Livermore, CA 94550, USA.

**Abstract:**

Solid state ionic approaches for modifying ion distributions in getter/oxide heterostructures offer exciting potentials to control material properties. Here we report a simple, scalable approach allowing for total control of the superconducting transition in optimally doped $YBa_2Cu_3O_{7-\delta}$ (YBCO) films via a chemically-driven ionic migration mechanism. Using a thin Gd capping layer of up to 20 nm deposited onto 100 nm thick epitaxial YBCO films, oxygen is found to leach from deep within the YBCO. Progressive reduction of the superconducting transition is observed, with complete suppression possible for a sufficiently thick Gd layer. These effects arise from the combined impact of redox-driven electron doping and modification of the YBCO microstructure due to oxygen migration and depletion. This work demonstrates an effective ionic control of superconductivity in oxides, an interface induced effect that goes well into the quasi-bulk regime, opening up possibilities for electric field manipulation.

**Keywords:** Superconductivity, Phase transitions, Ionic motion

**Introduction:**

Many of the properties of the high-temperature copper oxide superconductors are strongly influenced by charge doping.[1,2] The ability to control the doping level in these materials is not only essential for the development of experimental platforms for correlated electron physics, but also important for multifunctional device applications. While traditionally the doping level is fixed during synthesis via chemical substitution or post-growth annealing,[3–6] a number of recent approaches have demonstrated on-demand control of the doping level. By leveraging electrolytic double layer techniques, electrostatic gating experiments[7,8] on $RBa_2Cu_3O_{7-\delta}$ (R=Y, Nd) thin films have achieved control over the Cu-site doping level, which determines the dominant electronic order, by introducing oxygen vacancies into the film under electric fields. In these materials, $O^{2-}$ ions can migrate under the influence of an externally applied electric field to eventually escape through the film surface, resulting in the formation of oxygen vacancies. To maintain charge neutrality, electrons are returned to the Cu ions within the film, resulting in a reduction in average Cu valence and reducing the hole concentration of the oxide. These electrostatic, mostly interfacial effects, have profound impact on the electronic order in these materials, and point to the efficacy of using oxygen migration and vacancy formation to manipulate the properties of the cuprate superconductors.[9,10]

Recently, another solid state approach of ionic control of interface magnetism has been demonstrated in a number of systems,[11–18] utilizing the oxygen ion/vacancy transport across the metal/oxide interface. This electrochemical approach is different from the electrostatic approach which is usually confined to the electrolyte/material interface.[19] It enables highly efficient, nonvolatile control of essentially all magnetic functionalities via ion / vacancy migration, including anisotropy,[13,14,20] magnetization,[17] exchange,[16] etc. For example, we have shown effective manipulation of ionic distributions in oxide thin films by using a getter Gd capping layer.[15–18] Leveraging the reactivity of Gd, these capping layers can extract oxygen from an adjacent oxide film, with the level of oxygen depletion controlled by the thickness of the Gd and the ion mobility in the oxide, often at room-temperature. Interestingly, this ionic approach offers the possibility to manipulate the material properties into the bulk of the oxide films, well beyond the screening length of the metal/oxide interface region.[17,18,21,22] Control of the oxygen composition is particularly crucial in superconducting oxides, most of which require some amount of electron or hole doping, often achieved by introducing oxygen vacancies/interstitials.[4,23,24]

In this work, we demonstrate interfacial-redox-induced tuning of superconductivity throughout the entire thickness of 100 nm thick $YBa_2Cu_3O_{7-\delta}$ (YBCO) films. YBCO is a prototypical example of the high-$T_c$ (critical temperature) cuprates, with crystal structure and electronic ordering sensitive to oxygen stoichiometry. Combined with its high ionic conductivity,[25,26] these properties make YBCO an ideal candidate for explorations of ionic control. We find that Gd capping layers of up to 20 nm thickness deposited on 100 nm thick YBCO films can dramatically alter the oxygen distribution throughout the underlying film in the as-deposited state, without the need for a post-deposition annealing step. As the Gd layer thickness ($t_{Gd}$) is increased, the YBCO layer becomes progressively more oxygen deficient, demonstrating how appropriate tuning of $t_{Gd}$ can precisely control the remaining oxygen content of the underlayer. The extraction of oxygen from the YBCO suppresses the $T_c$ and broadens the superconducting transition. Differences between the superconducting transitions observed in resistivity and magnetometry suggest the formation of a percolative network of oxygen deficient regions interspersed within the nominally stoichiometric YBCO. Both the electron doping and structural changes induced by Gd-driven oxygen migration contribute to the suppression of superconductivity. These results demonstrate an effective solid-state ionic means to tailor superconductivity in cuprates and other systems, including the potential to use an electric field to tune superconductivity.

**Results:**

Commercially available epitaxial films of pulsed-laser-deposited *c*-axis oriented $YBa_2Cu_3O_{7-\delta}$ (100 nm) grown on (001)-oriented $SrTiO_3$ (STO) substrates were used for this study.[27] The high-symmetry STO substrate (cubic, lattice parameter of 3.905 Å) is known to yield YBCO films with 90° twinned domains,[28] with up to 2% tensile strain exerted at the interface due to lattice mismatch[29,30] (bulk lattice parameters of YBCO are $a$ = 3.827 Å, $b$ = 3.893 Å, and $c$ = 11.699 Å).[31] While the strain is known to suppress $T_c$, the films are expected to relax away from the substrate/film interface[32,33] and show near bulk-like superconducting behavior.[30] The films were then sputter coated with Gd layers of varying thickness ($t_{Gd}$ = 3 nm, 7 nm, 20 nm) and an Au (5 nm) protective cap, with one witness sample of YBCO/STO kept in the as-grown state for comparison.

An X-ray diffraction (XRD) scan of the as-grown film (Fig. 1A) shows only the (00$L$) family of peaks, corresponding to a lattice parameter of $c$ =11.678 Å, similar to the bulk value of 11.699 Å.[31,34] As $t_{Gd}$ is increased, a monotonic shift towards lower $2\theta$ is observed in the YBCO (00$L$) diffraction peaks, indicating expansion in the *c*-axis, similar to those previously reported in other perovskite systems[35] as a signature of oxygen depletion. The full width at half maximum (FWHM) of the diffraction peak also increases with $t_{Gd}$. In the as-grown sample, the (002) peak width is 0.08°, corrected for the instrument width, which matches the finite size broadening from the film thickness. In Gd-capped samples, the corresponding FWHM is 0.05°, 0.13° and 0.59° for the $t_{Gd}$ = 3nm, 7nm and 20nm sample, respectively. This noticeable broadening is manifestation of the spread in *c*-lattice parameter as oxygen is depleted from different depths of the YBCO, as well as a potential reduction in crystallite size. For the $t_{Gd} = 20$ nm sample, a significant shift and broadening of the (00*L*) peaks is observed, indicating further *c*-axis expansion as well as modification of the crystalline structure. The much broader peak width indicates a greater variation in the out-of-plane lattice parameter, indicating significant changes in the film crystallinity.

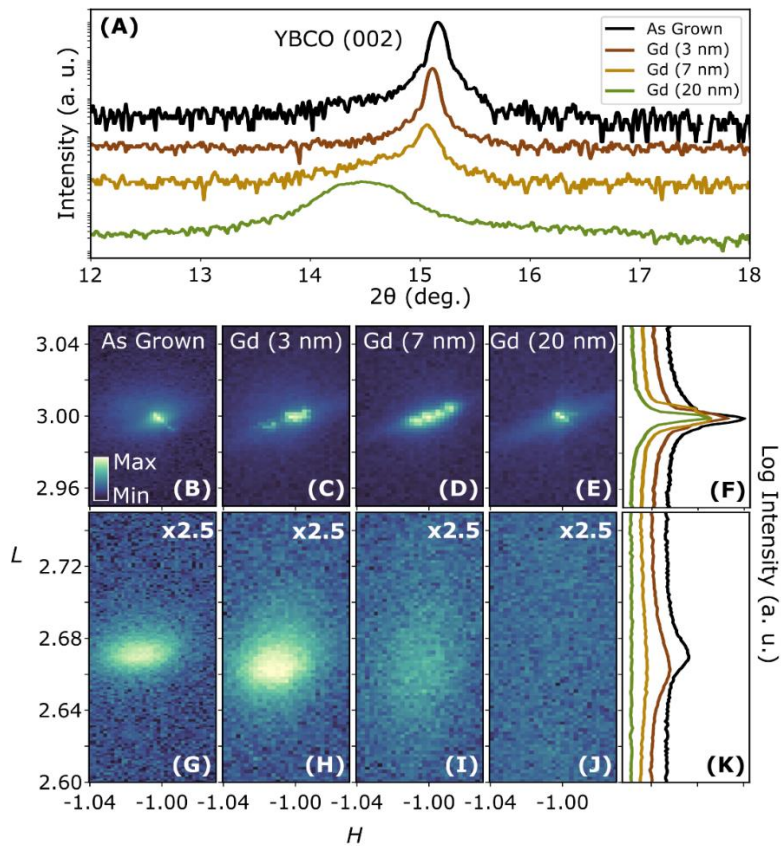

**Figure 1.** (A) X-ray diffraction pattern near the (002) YBCO peak measured with Cu $K\alpha_1$ radiation. Reciprocal space maps showing (B-E) the ($\bar{1}$03) STO substrate peak, (G-J) YBCO peak, indexed relative to the STO lattice parameters, and (F, K) Integrated RSMs along the *L*-index for each sample.

In addition to the out-of-plane direction, the in-plane structure of the films was probed using reciprocal space maps (RSMs) taken near the $(\bar{1}03)$ STO substrate reflection (Figs. 1B-E, G-J). Coordinates of the RSMs identify the *H* (*x*-axis) and *L* (*y*-axis) Miller indices relative to the STO substrate lattice parameters. When grown on cubic STO substrates, nominally orthorhombic YBCO is expected to form a twinned crystal structure, resulting in a splitting of the in-plane Bragg reflections.[29] For each sample, the RSM shows a bright, sharp substrate peak (Figs. 1B-E), while the streaking of the substrate peak in the $t_{Gd} = 3$ nm and 7 nm samples is likely due to some inhomogeneities in the substrate. Each of the $t_{Gd} = 0$ nm, 3 nm, and 7 nm samples also shows a lower-intensity, broadened YBCO peak near $H = -1.01, L = 2.67$; for $t_{Gd} = 20$ nm this peak is below our detection limit (Fig. 1G-J). The RSMs integrated along the *L*-index show the evolution of the substrate and film peaks across the different samples (Figs. 1F and 1K, respectively). Owing to the nearly identical in-plane lattice parameters, two closely spaced Bragg reflections are expected from the $(\bar{1}08)$ and $(0\bar{1}8)$ YBCO crystal planes; however, only one broad peak is observed in the as-grown YBCO (Fig. 1G). The alignment of the peak with the *H*= -1 coordinate indicates the epitaxial growth of the YBCO on the STO – again under tensile strain; however, the center of this peak is displaced towards more-negative values of *H* than the $(\bar{1}03)$ STO peak, indicating that these films are somewhat relaxed in-plane away from the interface, before Gd capping layers are deposited. As $t_{Gd}$ is increased, the YBCO film peak broadens along the (00*L*) direction and shifts to lower *L*, consistent with an expansion in the *c*-axis lattice parameter discussed above.

The effects of Gd deposition on the YBCO superconducting properties were investigated by magnetometry, as well as resistivity measurements in the van der Paw geometry. For magnetometry measurements, each sample was zero-field cooled to 5 K and measured on heating from 5 K to 100 K in a 1 mT magnetic field. Temperature dependence of the magnetization of the as-grown YBCO film shows a sharp superconducting transition at $T_c \approx 84$ K (Fig. 2), which is typical of YBCO films grown on STO substrates.[30] Below $T_c$, the magnetic flux is expelled from the superconducting YBCO due to the Meissner effect. Accompanying the magnetic transition is a precipitous drop in resistivity to zero, further confirming the superconducting transition. With increasing Gd capping layer thickness, the Meissner effect shows a reduction of $T_c$ to 62 K for $t_{Gd} = 3$ nm and 36 K for $t_{Gd} = 7$ nm, along with a broadened transition, and a complete suppression of superconductivity for $t_{Gd} = 20$ nm. In contrast, resistivity measurements initially show only a small $T_c$ reduction from 78 K in the as grown state to 76 K for $t_{Gd} = 3$ nm. However, for $t_{Gd} = 7$ nm and 20 nm the resistive superconducting transition is completely suppressed, with no apparent transition down to the lowest measured temperature (5 K). Note that no appreciable ferromagnetic signal was measured in any of the Gd-capped samples in 5 K-100 K, indicating that the Gd layer has been oxidized.

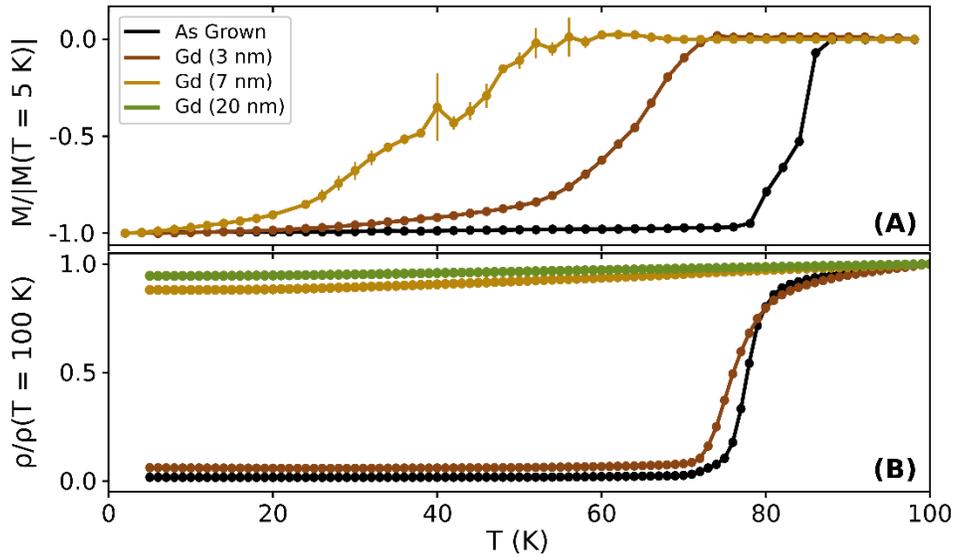

**Figure 2.** Superconducting transition measured by (A) normalized magnetization and (B) resistivity from 5 K to 100 K. The Gd (20 nm) sample has been omitted from (A), as no appreciable magnetic moment was detected in the temperature range measured (to within experimental noise).

While transport measurement probes a contiguous superconducting path, magnetometry is sensitive to the Meissner effect with contributions from the entire sample. The different temperature dependence observed in magnetometry and resistivity therefore reflects the inhomogeneities in the YBCO film, with magnetometry probing superconducting regions with any $T_c$, and resistivity measurement being most sensitive to regions of the film with the highest $T_c$ – so long as a superconducting path is maintained. The observed transitions can be explained by the presence of an oxygen-deficient phase with lower $T_c$, stabilized alongside the optimally doped YBCO phase. For the $t_{Gd} = 3$ nm sample much of the film possesses the as-deposited YBCO structure and the optimal stoichiometry of $YBa_2Cu_3O_{7-\delta}$; thus a contiguous path exists through the optimally doped phase, and the resistive transition is observed close to the $T_c$ of the as-grown film, even though the sample-averaged Meissner effect shows a much lower $T_c$ with a more gradual transition. For the $t_{Gd} = 7$ nm sample, an even broader transition is observed in the magnetometry, but no complete suppression of resistivity is observed. Therefore, this sample contains some oxygen deficient regions which are still superconducting at lower $T_c$, as seen in magnetometry, but these phases are below the percolation threshold to form a contiguous superconducting path, as evidenced by the absence of a resistive transition. Finally, for the thickest $t_{Gd} = 20$ nm sample the absence of a transition in either the resistance or magnetization suggests few, if any, regions undergo superconducting transition in the sample. The distinctly different superconducting properties of the oxygen deficient phases may be the result of structural changes or electron doping, both of which are consequences of the oxygen leaching.

The Gd capping layer is expected to extract oxygen from the YBCO, resulting in the structural changes that were observed in XRD patterns. Cross-sectional high-angle annular dark field (HAADF) images of the as-grown film at the STO/YBCO interface obtained in an aberration-corrected scanning transmission electron microscope (STEM) show flat, epitaxial *c*-axis oriented YBCO growth consistent with XRD patterns (Fig. 3A). The layered structure of $YBa_2Cu_3O_{7-\delta}$ is well resolved, as illustrated in a zoomed-in view of the STO/YBCO interface shown in Fig. 3A inset (right half of the inset). The CuO chains are clearly manifested as darker contrast regions located between perovskite blocks consisting of BaO-$CuO_2$-Y-$CuO_2$-BaO, which exhibit a brighter contrast due to the higher average Z-number. A number of CuO stacking faults, which disrupt the YBCO epitaxy, are found in the imaged region of Fig. 3A, highlighted by red dashed lines. They may appear due to the splitting of a single CuO chain plane or by insertion of an extra CuO chain layer, which cannot be conclusively distinguished here. In the Gd (3 nm)/YBCO sample, the lateral extent of these stacking faults increases as the film becomes increasingly oxygen deficient, as shown in Fig. 3B. An example is illustrated in Fig. 3B inset, where a CuO stacking fault consisting of multiple CuO chain layers is seen as a wider, dark-contrast region in the middle of bright contrast perovskite blocks. The number of stacking faults increases even further for the Gd (20 nm)/YBCO sample, but these defects are more fragmented than those observed for $t_{Gd} = 3$ nm, as shown in Fig. 3C (see Supporting Information for additional microscopy results). Effectively these defects "carve out" the epitaxial YBCO film into smaller crystallites, consistent with the aforementioned XRD peak broadening. Examples of similar stacking faults have been previously reported in pulsed laser deposited films of YBCO,[36–38] and attributed to the limited atomic diffusion range under typical PLD conditions. Double- or multiple-CuO stacking faults form different microstructural phases than optimally doped $YBa_2Cu_3O_{7-\delta}$ (Y-123, with numbers denoting cation stoichiometry).[39] These defects directly alter the CuO chains, which act as charge reservoirs for superconductivity in the YBCO system. Accordingly, these types of defects affect the superconducting transition, and result in lowered critical temperatures as compared to Y-123.[40–42] These stacking faults point to the aggressive nature of the oxygen leaching effect, as the YBCO layers become increasingly disrupted when oxygen is removed from deep within the film, and suggest that changes to the microstructure are directly correlated with the suppression of superconductivity.

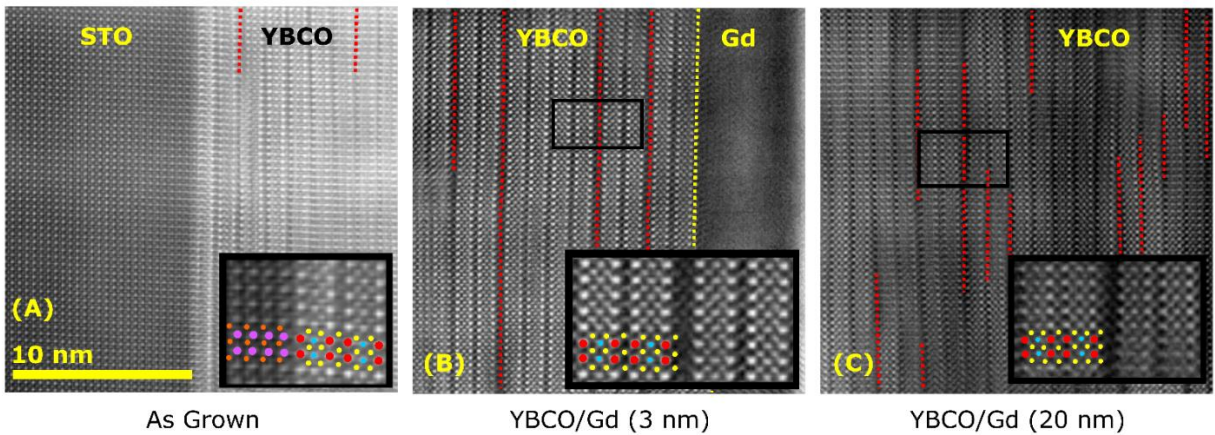

**Figure 3.** Cross-sectional HAADF-STEM images of (A) the as-grown STO/YBCO interface with a magnified view of the substrate/film interface shown in the inset (which is outside the field of view of the main image). Some spatial drift is visible. The crystal structure is depicted with Cu, Ba, Y, Sr, and Ti atoms highlighted in yellow, red, blue, violet, and orange, respectively. (B) The YBCO/Gd interface of the Gd (3 nm) sample with a magnified view of the region highlighted by the black box shown in the inset. (C) Center of the YBCO layer in the Gd (20 nm) film, with a magnified view of the region indicated by the black box shown in the inset. Both insets in (B) and (C) contain vertically oriented CuO stacking faults in the middle (double and triple CuO chain layers), similar to those in (A-C) highlighted by dashed red lines.

The extraction of oxygen was further probed by polarized neutron reflectometry (PNR) at 6 K, which provides a depth-resolved mapping of the nuclear scattering centers within the film (neutron reflectivity data are shown in Supporting Information). The converged depth profiles show that the nuclear scattering length density (SLD, $\rho_N$) of the as-grown film is around the bulk value for YBa$_2$Cu$_3$O$_7$ of $4.7 \times 10^{-4}$ nm$^{-2}$ (Fig. 4A). With increasing $t_{Gd}$, the YBCO layer increases in thickness, in qualitative agreement with the unit cell expansion observed in XRD. Commensurate with the progressive increase in $t_{Gd}$, the amplitude of the Kiessig fringes in the reflectivity decreases, indicating a reduction of contrast of the YBCO vs. the STO, which results from their nuclear SLDs converging. Indeed, the fits confirm that as $t_{Gd}$ increases, $\rho_N$ decreases, particularly near the STO/YBCO interface (Figs. 4B-4C). The decrease in $\rho_N$ near the substrate in the $t_{Gd}$ = 3 nm sample indicates that oxygen vacancies introduced at the YBCO/Gd interface migrate through the film and accumulate at the bottom STO/YBCO interface. This is likely a mechanism to relieve the tensile strain in the as-grown film. Specifically, the STO lattice parameter is ≈2% larger than the YBCO when it is stoichiometrically balanced. The X-ray diffraction indicates that removing oxygen causes the lattice parameter to expand. For the $t_{Gd}$ = 3 nm sample, the initial accumulation of oxygen vacancies at the STO/YBCO interface causes the YBCO lattice parameter to expand, better-matching the STO lattice parameter and thus reducing the tensile strain. However, as more vacancies are introduced in the $t_{Gd}$=7 nm sample, further oxygen vacancies at this STO/YBCO interface would generate a compressive strain – as the YBCO lattice parameter continues to expand beyond the STO lattice parameter. To avoid this new strain energy, oxygen vacancies in the $t_{Gd}$= 7 nm sample tend to accumulate at the top YBCO/Gd interface. In addition to the real part of the nuclear profile, the Gd layer is explicitly identifiable by the imaginary component of its SLD, which corresponds to neutron absorption. Since Gd is the only significant neutron absorber present in this system, the location of the imaginary component of the nuclear SLD rules out the possibility of significant YBCO/Gd interdiffusion. This fact also contributed to particularly poor statistics when measuring the $t_{Gd}$ = 20 nm sample – large neutron absorption severely attenuated the reflectometry. Furthermore, no appreciable magnetic SLD was measured (Supporting Information, Fig. S1), indicating that the Gd layers are no longer magnetic, in agreement with magnetometry measurements.

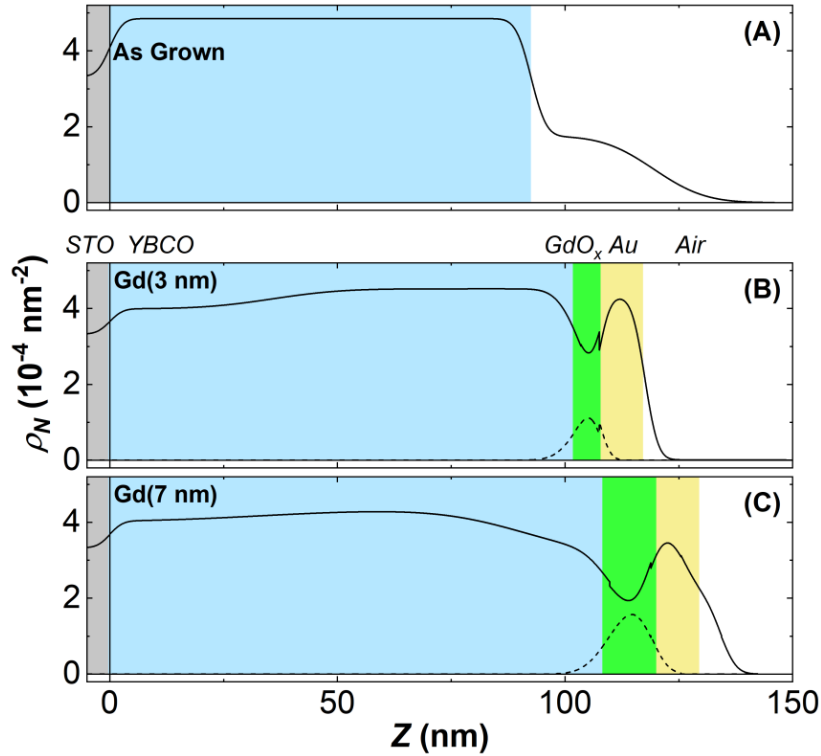

**Figure 4.** Real and imaginary components of the nuclear SLDs as a function of depth $Z$ through the (A) as grown, (B) Gd (3 nm), and (C) Gd (7 nm) samples as measured by PNR. Grey region ($Z < 0$) corresponds to the substrate, with the approximate location of the YBCO, Gd, and Au layers given by the shaded regions for ($Z > 0$).

The PNR results are consistent with the extraction of oxygen from the YBCO, and fluorescence yield (FY) XAS measurements performed at the Cu $L_{2,3}$-edges directly confirm a change in the Cu valence induced by the oxygen extraction. The XAS results show a shift in the absorption resonance to lower energies with increasing $t_{Gd}$ (Fig. 5). Similar spectral shifts reported in other oxygen-deficient perovskite systems,[43–50] have been attributed to a decrease in the average Cu valence, the result of electrons returning to the Cu ions as oxygen is leached from the film. The shoulder around 933 eV in the as-grown YBCO spectra (marked with a black arrow), a feature characteristic of lower-valence ligand states present in CuO chains,[8,51,52] is suppressed for YBCO coated with increasingly thicker Gd, confirming the loss of oxygen within the chains. A second resonance associated with the $Cu^{1+}$ valence state emerges at 934 eV for $t_{Gd} = 7$ nm and 20 nm; this absorption peak was previously reported in oxygen deficient bulk YBCO, and expected to be absent for samples with optimal oxygen stoichiometry.[51] While reduction is known to occur at metal/YBCO interfaces,[43–48] it was reported to be limited to the surface region (within 5 nm) due to oxygen migration kinetics. The bulk sensitivity of FY measurements infers that the observed oxygen depletion comes from ionic migration from deep within the film, despite the interfacial origin of the leaching effect. The high ionic conductivity required for such long-range oxygen migration at room temperature is in agreement with previous reports.[25,26] The known sensitivity of the YBCO superconductivity to the oxygen stoichiometry suggests that this ionic approach may therefore be used as an effective means to control the superconducting transition, potentially under an electric field.

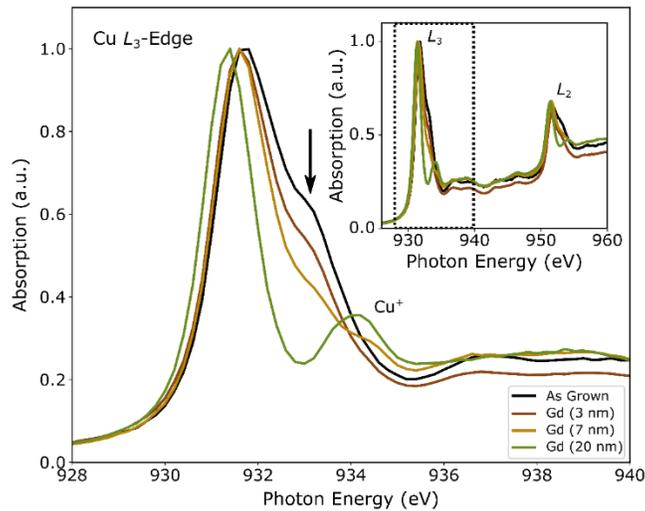

**Figure 5.** Close-up view of the normalized absorption spectra as a function of incident X-ray photon energy near the Cu $L_3$-edge, as measured in fluorescence yield mode. The full spectral range measured including the Cu $L_2$-edge is shown in the inset, with the dashed box corresponding to the close-up view. The arrow marks the shoulder at 933 eV.

**Discussion:**

The removal of oxygen from YBCO by the Gd capping layer can be understood primarily by considering the change in Gibbs free energy associated with oxidizing the Gd and reducing the YBCO. At room temperature (300 K), the Gibbs energy of formation $\Delta G$ for the constituent $Y_2O_3$, BaO, $BaO_2$, and CuO oxides are -1817 kJ/mol, -520 kJ/mol, -588 kJ/mol, and -130 kJ/mol, respectively,[53,54] with the enthalpy of formation $\Delta H$ = -86 kJ/mol for $0.5Y_2O_3 + 1.5BaO + 0.5BaO_2 + 3CuO \rightarrow YBa_2Cu_3O_7$.[55] Since the entropy of formation of the $YBa_2Cu_3O_7$ compound from the constituent oxides is expected to be small,[56] the total Gibbs free energy of formation from the base elements can be estimated by the sum of these energies, $\Delta G_{YBCO}$ = -2458 kJ/mol. Given the large Gibbs energy of formation of $Gd_2O_3$, $\Delta G_{Gd_2O_3}$ = -1730 kJ/mol, the net change in Gibbs free energy for the reaction of interest,

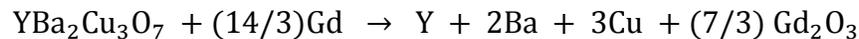
$$YBa_2Cu_3O_7 + (14/3)Gd \rightarrow Y + 2Ba + 3Cu + (7/3)\,Gd_2O_3$$

is estimated to be $\Delta G$ = -1578 kJ/mol, indicating that the Gd cap can be expected to spontaneously reduce the YBCO. The formation of other intermediate oxide phases, including $Y_2Ba_4Cu_6O_{13}$ and $YBa_2Cu_3O_6$, is expected to occur at even lower energies, and thus may occur during the reduction process. The formation of these minority phases should be manifested in the Cu valence reduction, which is consistent with the XAS results; the formation of $YBa_2Cu_3O_6$ is identifiable by the $Cu^{1+}$ valence, which indeed is observed in the XAS. In addition to formation energies, the ability to extract oxygen from throughout the large thickness of the YBCO is determined by the electron work function,[57] which is characteristically low in Gd. This is in contrast to previous studies of capped YBCO layers where oxygen extraction was found to be limited to the surface, using capping layers with larger work functions and lower formation energies.[43–48]

Two different mechanisms may contribute to the suppression of superconductivity. The first mechanism is electron doping, as observed in XAS, where the more oxygen is stripped away by the adjacent Gd capping layer, the more the Cu valence state decreases. This electron doping effectively shifts the YBCO film from the optimally doped state, with maximum $T_c$, back towards the electron-doped region of the global phase diagram, with a corresponding decrease in $T_c$. Magnetometry shows that for $t_{Gd}$=3 nm and 7 nm, the superconducting transition broadens across a range of temperatures, indicating regions with a distribution of $T_c$ coexist within the same film. The second mechanism involves the disruption to the film microstructure. The crystal structure of optimally doped YBCO contains both $CuO_2$ planes – where superconductivity resides – and CuO chains, which act as doping centers for the planes. As oxygen is removed from the film, the chains become progressively more oxygen deficient,[26] leading to more defects present in the YBCO film, as observed by STEM. The presence of double- and multiple-CuO stacking faults directly modifies these doping centers, locally forming separate phases with lower values of $T_c$. In addition to these extended defects, YBCO is also known to be susceptible to point defects or disorder, owing to its short superconducting coherence length ($\xi \approx 20$ Å). Studies of ion-irradiation-induced disorder have shown that with increasing defect density, the electron mean free path is reduced, suppressing $T_c$.[58–61] In the present study, Gd capping may also reduce $T_c$ by suppressing the electron mean free path, as the density of CuO stacking faults as well as disorder arising from oxygen migration is expected to correlate with $t_{Gd}$. Finally, it is worth noting that the ferromagnetism of Gd is absent in these Gd/YBCO samples, as confirmed by both PNR studies (Supporting Information, Fig. S1) and magnetometry (Supporting Information, Fig. S2), and is not expected to contribute to the suppression of superconductivity.

In summary, thin Gd capping layers deposited onto optimally doped YBCO thin films have been shown to remove oxygen from deep within the underlying film via an interfacial redox reaction, with the extent dependent on capping layer thickness. This redox-induced oxygen migration, although initiated at the Gd/YBCO interface, results in a percolating oxygen deficient phase throughout the entire film thickness that does not support superconductivity. As measured in magnetometry and resistivity, the superconducting transition temperature is significantly reduced with increasing Gd thickness, and for a sufficiently thick capping layer of up to 20 nm, the transition is completely suppressed. Spectroscopic measurements reveal the loss of oxygen with the CuO chains as well as a decrease in the average Cu valence, indicating that oxygen leaching effectively acts to electron dope the YBCO and thus decreases $T_c$. STEM images reveal the presence of defects associated with separate, oxygen-deficient phases, suggesting that both changes to the microstructure as well as a reduction in hole doping level may play roles in the suppression of superconductivity. Remarkably, the changes to the superconducting properties throughout the entire 100 nm thick YBCO films are induced by the migration of oxygen towards the YBCO/Gd interface, extending the viability of ionic control of superconductivity to the quasi-bulk regime. As ionic transport can be easily controlled by a bias voltage, our findings also show potential in electric field control of superconductivity in getter/oxide type of heterostructures.

**Supporting Information**. The following files are available free of charge. Additional experimental details on neutron reflectivity, magnetometry, and electron microscopy studies of the sample microstructure (PDF).

**AUTHOR INFORMATION**

**Corresponding Author**

* Corresponding author: Kai.Liu@georgetown.edu

**Author Contributions**

P.D.M., D.A.G. and K.L. designed and coordinated the project. P.D.M. synthesized the samples, carried out structural analysis, and wrote a first draft of the manuscript. D.A.G., A.J.G., B.J.K, W.L.N.C.L. and J.A.B. performed PNR studies. D.A.G., A.J.G. and E. A. carried out XAS measurements. D. H-M. and M.V. carried out electron microscopy studies. V. T. measured magnetometry. Z.E.B., J.R.J., and R. Z. carried out transport measurements. R.V.C. and Y.T. assisted with sample synthesis and RSM studies. All authors contributed to data analysis and manuscript revision. ‡P.D.M. and D.A.G. contributed equally.

**Notes**

The authors declare no competing financial interests.

**Film growth and characterization.** Commercially available 100 nm thick YBCO films grown on STO substrates were purchased from MTI Corporation for this experiment. The films, which were packaged and kept in a vacuum-sealed box prior to Gd deposition, were exposed to atmosphere for < 1 h before being transferred to a high vacuum environment at UCD. The YBCO films were subsequently sputter-coated at room temperature with Gd (3 nm, 7 nm, 20 nm) and a Au (5 nm) protective capping layer using Ar gas at 0.67 Pa working pressure in a chamber with a base pressure in the $10^{-6}$ Pa range. XRD characterization, including both $\theta - 2\theta$ scans and reciprocal space maps, was performed on a X-ray diffractometer equipped with parallel beam optics and Cu $K\alpha_1$ monochromator. Polarized neutron reflectometry was measured at the NIST Center for Neutron Research on the PBR and MAGIK beamlines. The experiments used 4.75 Å and 5 Å neutrons respectively, and were carried out at a temperature of 6 K. Fitting of the PNR data was performed using the Refl1d software package, following a Markov-chain Monte Carlo fitting algorithm.[62] The calculated SLD was determined by the calculating the sum of the volume-scaled atomic scattering lengths. XAS measurements were performed at the Advanced Light Source on beamline 4.0.2 at room temperature in a grazing incidence (30°) geometry. Both fluorescence and electron yield (EY) modes were measured, but due to the capping layers no appreciable signal was measured in the EY mode, thus only FY data are shown. Magnetometry measurements were performed by first cooling the samples to 5 K in zero field; a field of 1 mT was then applied, and the magnetic moment was recorded as the sample was warmed up to 100 K. To reduce stray magnetic fields and trapped flux, the superconducting magnet was driven into the normal state before each measurement. Resistivity was measured using a four-contact van der Pauw geometry. The excitation current used was 100 μA at a frequency of 173 Hz. The transition temperatures in the magnetometry and resistance measurements were acquired by taking the midpoint of the transitions. Cross-section specimens for electron microscopy were prepared by grinding, polishing and ion milling with a final 0.5 kV cleaning. STEM analyses were carried out in a JEOL ARM200cF equipped with spherical aberration corrector and working with an acceleration voltage of 200 kV at the National Center for Electron Microscopy at University Complutense of Madrid, Spain.

**Acknowledgements.** This work has been supported by the NSF (DMR-1610060 and ECCS-1611424). Work at GU has been supported in part by SMART, one of seven centers of nCORE, a Semiconductor Research Corporation program, sponsored by National Institute of Standards and Technology (NIST). Work at LLNL has been supported by the DOE (DE-AC52-07NA27344) and NSF (DMR-1609855). Use of the Advanced Light Source was supported by DOE Office of Science User Facility under Contract No. DE-AC02-05CH11231. Work at UCM supported by MINECO-FEDER grants MAT2015-66888-C3-3-R and MAT2017-89599-R, and also by the TALENTO program, Comunidad de Madrid.